\documentclass[fleqn,10pt]{SelfArx} 

\definecolor{color1}{RGB}{0,0,90} 
\definecolor{color2}{RGB}{0,20,20} 

\usepackage{hyperref} 
\hypersetup{hidelinks,colorlinks,breaklinks=true,urlcolor=color2,citecolor=color1,linkcolor=color1,bookmarksopen=false,pdftitle={Title},pdfauthor={Author},urlcolor=blue}

\usepackage{graphicx}
\usepackage[square]{natbib}
\usepackage{amsmath}
\usepackage{multirow}
\usepackage{subfigure}
\usepackage{pdflscape}

\newcommand{\n}[1]{\mathrm{#1}}

\JournalInfo{Published in Journal of Magnetism and Magnetic Materials, Vol. 476, 417-422, 2019} 
\Archive{\href{https://doi.org/10.1016/j.jmmm.2019.01.005}{DOI: 10.1016/j.jmmm.2019.01.005}} 

\PaperTitle{The demagnetization factor for randomly packed spheroidal particles} 

\Authors{R. Bj\o{}rk$^1$ and Z. Zhou$^2$} 
\affiliation{\textsuperscript{1}\textit{Department of Energy Conversion and Storage, Technical University of Denmark - DTU, Frederiksborgvej 399, DK-4000 Roskilde, Denmark}} 
\affiliation{\textsuperscript{2}\textit{Laboratory for Simulation and Modelling of Particulate Systems, Department of Chemical Engineering, Monash University, Clayton, VIC 3800, Australia}} 
\affiliation{*\textbf{Corresponding author}: rabj@dtu.dk} 

\newcommand{\keywordname}{Keywords} 

\Abstract{We investigate if the demagnetization factor for a randomly packed powder of magnetic spheroidal particles depend on the shape of the spheroidal particles and what the internal variation in magnetization is within such a powder. A spheroid is an ellipsoid of revolution, i.e. an ellipsoid with two semi-major axis being equal. The demagnetization factor is calculated as function of particle aspect ratio using two independent numerical models for several different packings, and assuming a relative permeability of 2. The calculated demagnetization factor is shown to depend on particle aspect ratio, not because of direct magnetic interaction but because the particle packing depend on the aspect ratio of the particles. The relative standard deviation of the magnetization across the powder was 3\%-8\%, increasing as the particle shape deviates from spherical, while the relative standard deviation within each particle was relatively constant around 5\%.}

\begin{document}

\flushbottom 

\maketitle 


\thispagestyle{empty} 

\section{Introduction}
The magnetic response of a sample to an external magnetic field is important in a wide range of fields, from nanomagnetic containers for medicine to magnetic refrigeration. In general, the magnetic properties of a sample when subjected to an external magnetic field depend on the magnetic properties of its constituent magnetic particles. If the sample is solid, certain sample geometries, such as an ellipsoid or an infinite cylinder, will have a uniform magnetization in a uniform external field \cite{Osborn_1945}.

However, most magnetic samples do not have a geometry that results in uniform magnetization, nor are most samples necessarily solid. When subjecting a magnetic sample to an external field, the sample generates a magnetic field that causes a self-interaction with the magnetic material of the sample. This field is termed the demagnetization field, and the internal field within the sample can be expressed with regards to this self-interaction as
\begin{eqnarray}\label{Eq.Demag_def}
\mathbf{H}=\mathbf{H}_\textrm{appl}-\mathbb{N}\cdot{}\mathbf{M}
\end{eqnarray}
where $\mathbf{H}$ is the magnetic field, $\mathbf{H}_\textrm{appl}$ is the externally applied field, $\mathbf{M}$ is the magnetization and $\mathbb{N}$ is the demagnetization tensor.
To express the above equation as a scalar equation the demagnetization tensor can be replaced with an average demagnetization factor, $N$. For prism-shaped magnetic particles with varying aspect ratios from 0.05 to 10, this scalar approximation has been shown to deviate only 1\% compared to the full demagnetization tensor \cite{Smith_2010}. However, for thin disc-shaped samples the deviation of the average demagnetization factor from the true result can be as much as  10–20\%, depending on the coercivity and state function of the sample and the geometrical properties of the disc \cite{Brug_1985}.

For some sample geometries the demagnetization tensor can be analytically computed, e.g. for a prism \cite{Joseph_1965,Smith_2010}. The average demagnetization factor can also be analytically calculated for a range of geometries, such as cuboids \cite{Aharoni_1998} and cylinders \cite{Joseph_1966}. These demagnetization factors only apply to solid samples, and not samples consisting of magnetic powders, which are also of great interest scientifically and practically. If the powder is regularly packed, such as  simple-cubic, body-centered-cubic and face-centered-cubic packed spheres, the demagnetization factor is known \cite{How_1991}. This is also the case for aggregates of particles in a matrix \cite{Skomski_2007}, various two-dimensional arrays of magnetic particles \cite{Martinez_Huerta_2013} and cylinders \cite{Chen_2006}.

If the powder is randomly packed, and consists of spherical particles both Breit \cite{Breit_1922} and Bleany \cite{Bleaney_1941} have shown that the demagnetization factor can be expressed as
\begin{eqnarray}\label{Eq.Demag}
N=\frac{1}{3}+f\left(D-\frac{1}{3}\right)
\end{eqnarray}
Here $D$ is the demagnetization factor of the geometrical shape of the powder sample itself and $f$ is the relative density. The factors of $1/3$ in the equation ensure that the demagnetization factor is $1/3$ for any object that is symmetric with respect to the direction that the field is applied, such as a sphere or a cube with the field perpendicular to one of its faces. In deriving the above equation it is assumed that all particles are spheres and that these are uniformly magnetized. However, magnetic particles are seldom perfect spheres  \cite{Park_1997,Ott_2009} and the particles will not be uniformly magnetized within a packed sample \cite{Bjoerk_2013}.

For both spherical and non-spherical particles the magnetic properties of magnetic particles bonded as a sample has been studied for various volume fractions of the magnetic powder in numerous studies \cite{Le_Floch_1995, Walmsley_2000, Mattei_2003a, Anhalt_2007}. The detailed properties as function of volume fraction \cite{Mattei_2003b}, for complex values of the permeability \cite{Drnovsek_2008}, agglomerations \cite{Drnovsek_2008}, 2D geometries \cite{Mattei_2000}, conductive inclusions \cite{Drnovsek_2009}, coercivity \cite{Anhalt_2009}, particle size \cite{Anhalt_2008}, dynamic effects \cite{Chevalier_2001}, particle size distribution \cite{Bjoerk_2013}, particle shape (rods, cylinders, oblate spheroids) \cite{Mattei_1996} and particle permeability \cite{Bjoerk_2013} have all been studied. Recently, tomography scans of bonded soft magnetic particles have also been used to calculate the demagnetization factor \cite{Arzbacher_2015}, which was compared with computational packings as function of volume fraction with good agreement.

However, while all of these studies have focused on calculating the demagnetization factor, the variation in internal magnetization both among the magnetic particles and within each individual particle have only been considered for spherical particles \cite{Bjoerk_2013}. It is important for physical phenomena dependent on the internal magnetization, such as for example the magnetocaloric effect for which the temperature change usually depends on the internal magnetic field to the power of 2/3 \cite{Bjoerk_2010d}, that the internal magnetization within a given sample is known in detail. In magnetic refrigeration packed beds of particles, both spherical and irregular in shape have been modeled \cite{Tusek_2013,Lei_2017}  and also tested experimentally \cite{Navickaite_2017}, and for these it is important to know the correct internal magnetization.

Here we present modeling, using two different approaches, of packings of magnetic powders with spheroidal particles, which have often be used to approximate real magnetic particles \cite{Fan_2003,Yelon_1971}. The objective is to determine how the particle shape influence the magnetic properties of the powder packing, i.e. the demagnetization factor, and study the internal magnetization throughout the sample and within the individual particles. Also, it is not known if Eq. (\ref{Eq.Demag}) is correct for powders with non-spherical particle shapes. This is also investigated here.

\section{Modeling setup}
The demagnetization factor is investigated using different numerical frameworks. Note that a magnetostatic problem is scale invariant, and thus the length scale is arbitrary, as long as the particles considered are much larger than the magnetic domains. Thus the results here do not apply to packings of nanoparticles \cite{Skomski_2010}. By this we mean that the particles considered in this study all behave as macroscopic particles with regards to magnetization, etc., i.e. they are not dominated by the energy associated with domain walls and other micromagnetic effects.

The finite element framework (FEM) Comsol Multiphysics is used to solve the magnetostatic problem of calculating the demagnetization factor. This has also been used in previous studies \cite{Drnovsek_2008, Bjoerk_2013}. The equation solved in the FEM framework is the magnetic scalar potential equation
\begin{eqnarray}
-\nabla{}\cdot{}(\mu{}_{0}\mu{}_{r}\nabla{}V_\mathrm{m})=0~.\label{Eq.Numerical_Magnetism}
\end{eqnarray}
Here $\mu{}_{0}$ is the permeability of free space, $\mu{}_{r}$ is the relative permeability, which is assumed to be constant and isotropic, and $V_\mathrm{m}$ is the magnetic scalar potential. The magnetic field is then calculated as $-\nabla{}V_\mathrm{m} + \mathbf{H}_\textrm{appl} = \mathbf{H}$, where $\mathbf{H}_\textrm{appl}$ is the applied magnetic field. In all experiments, the applied magnetic field is along the $z$-axis and has a magnitude of $\mu_0\mathbf{H}_\textrm{appl}=1$ T. To indicate the field direction for the computed demagnetization factor, the demagnetization factor and the demagnetization factor of the geometrical sample shape are denoted $N_\n{z}$, and $D_\n{z}$, respectively.

The built-in Comsol Multiphysics solver \emph{Pardiso} which is a parallel sparse direct linear solver \citep{Schenk_2001,Schenk_2002} is used to solve the equation on the finite element mesh. The computational volume is chosen large enough that the boundaries of the simulation volume do not affect the calculations.

Furthermore, the demagnetization factor is also computed using an independent numerical framework that can calculate the magnetic field from magnetic particles with a shape for which the demagnetization tensor is known \cite{Smith_2010}. The field is calculated directly from the analytical expression of the demagnetization tensor. The only assumption is that each particle internally have a constant magnetization. The model has previously been used for prisms \cite{Nielsen_2017}, but has here been extended to spheroidal particles, based on the expression of the demagnetization tensor for spheroids given in Ref. \cite{Tejedor_1995}. This model is included here to investigate the often used simplifying assumption that the magnetic particles can be assumed to be individually uniformly magnetized.

The magnetic particles are assumed to consist of a linear magnetic material with a constant $\mu_r=2$. This is approximately the value for gadolinium at a magnetic field of 1 T \cite{Bjoerk_2010d}, but it is also appropriate to other ferromagnetic materials, as is shown in Appendix A. Note that the demagnetization factor for a linear material depends on the relative permeability \cite{Bjoerk_2013}. The volume fraction of particles, i.e. the relative density of the packing, is estimated using a Monte Carlo sampling method in the inner 50\% of the sample, in order to disregard surface effects.

\begin{figure}[t]
  \centering
  \includegraphics[width=1\columnwidth]{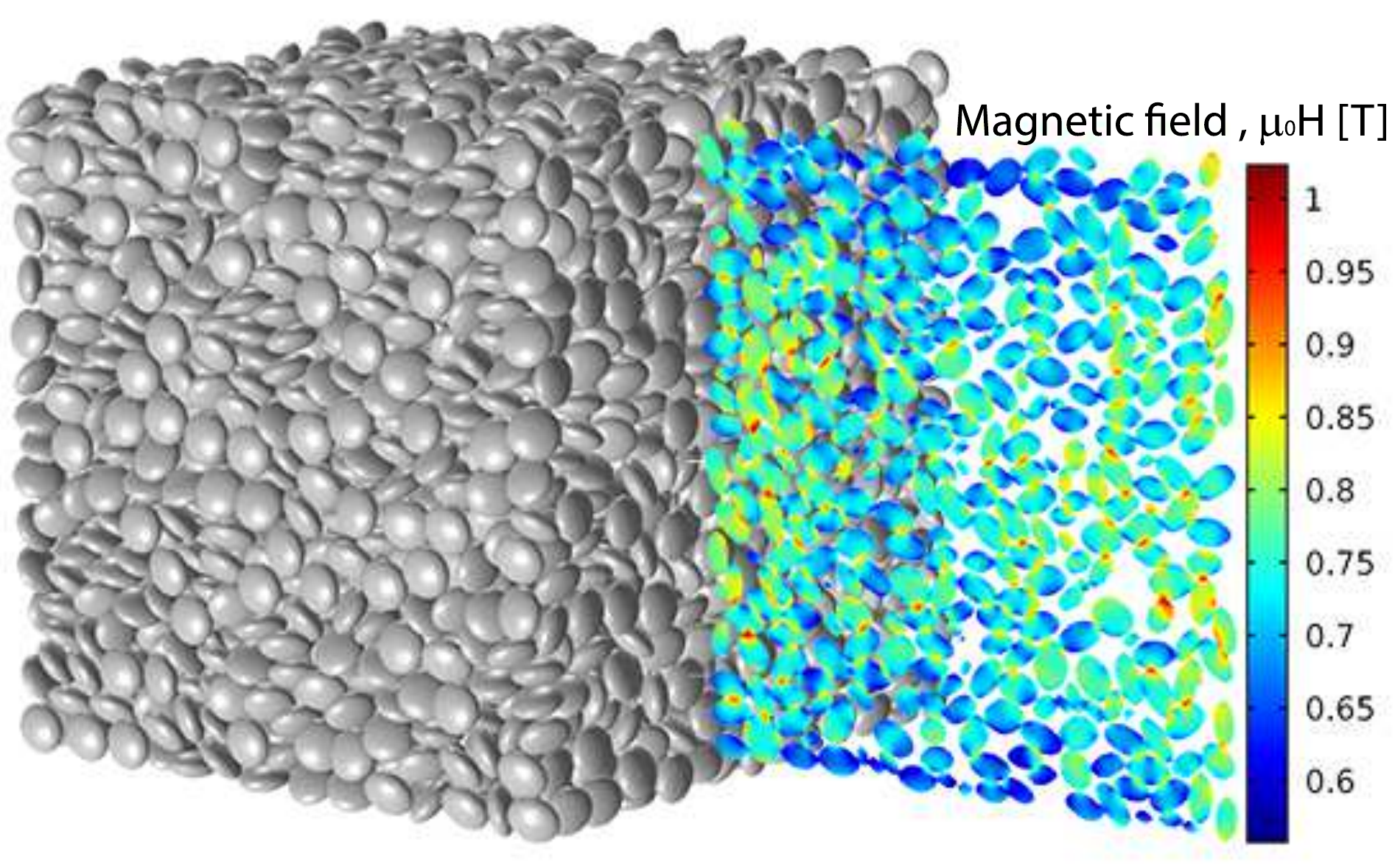}
  \caption{An illustration of a randomly packed powder of spheroids with a particle aspect ratio of 0.5. The computed internal magnetic field in the particles in a slice through the sample is shown next to the sample.}
    \label{Fig.Packing_illustration}
\end{figure}

\begin{figure}[h!]
  \centering
  \includegraphics[width=0.7\columnwidth]{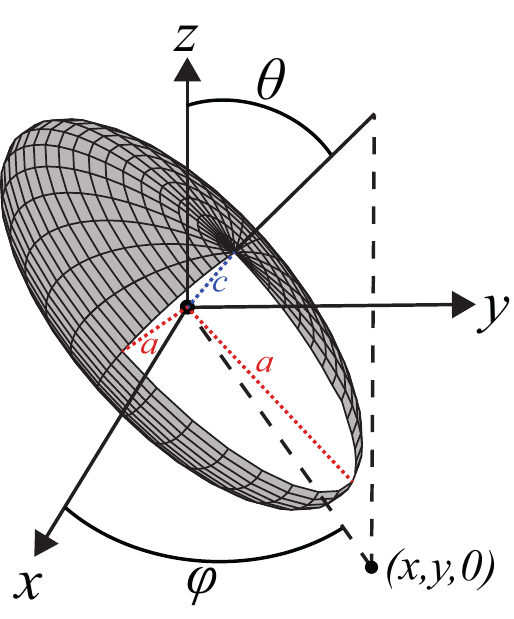}
  \caption{The semi-major axis, $a$ and $c$, and the spherical orientation angles for an oblate spheroid. In spherical coordinates, $\phi$ denoting the azimuthal angle and $\theta$ denoting the polar angle.}
    \label{Fig.Spheroid_illustration}
\end{figure}

The powder packing of the spheroidal particles were generated using the framework described in Ref. \cite{Zhou_2011}. In this modeling framework, particles of the desired shape are simulated being poured into a container and allowed to settle under gravity. Powders can generally be made under the influence of other other driving forces such as liquid solvents, surfactants, pressure, applied field, but gravity is arguably the simplest driving force to consider. When samples are made by pouring a magnetic powder into a sample holder, gravity is the driving force, and this situation is what is considered here. In the framework used to simulate the particle packing, the particles are simulated in full 3D and the particles cannot overlap, as opposed to a previous study \cite{Arzbacher_2015}. To reduce the parameter space, we consider spheroids, i.e. ellipsoids with two of the semi major axis equal to each other. Spheroids with an aspect ratio of 0.5, 0.75, 1, 1.5 and 2 are considered, as well as a packing with a mix of particles in equal proportions of these. The aspect ratio is the ratio between the two semi major axis of the spheroid, and thus both oblate ($<1$) and prolate ($>1$) particles are considered. Two differently shaped samples were considered; a cube sample and a short cylindrical sample. The latter has a diameter about 2.35 times the value of its height, and here the magnetic field is applied along the height direction. This ratio was chosen a priori to make the cylinder have a reasonable high demagnetization factor. The cube must have a demagnetization factor of 0.33, if the cube is homogeneous. This follows from the fact that the different face demagnetization factors must sum to one, i.e. that the trace of the demagnetization tensor is 1 \cite{Moskowitz_1966}. The short cylinder is chosen to explore the particle influence on a sample with a high demagnetization factor. The cube samples have on average $4713\pm144$ particles while the cylindrical samples have $2320\pm{}48$ particles. An illustration of a packing of spheroidal particles with an aspect ratio of 0.5 in a cube is shown in Fig. \ref{Fig.Packing_illustration}. For each particle aspect ratio, five different packings were generated in order to calculate a statistical uncertainty on the computed properties. A single oblate spheroid and the parameters that define this is illustrated in Fig. \ref{Fig.Spheroid_illustration}.

When a magnetic field is applied to a packing of non-spherical particles, these will possibly reorient and rearrange their position to align with the field. In this work, this reorientation and rearrangement effect is not taken into account as it is too computationally intensive to solve for as a static magnetic framework can not longer be used. Practically this means that the packings consider here are clamped tightly or the particles are bonded in a non-magnetic medium e.g. epoxy, which is also done experimentally \cite{Navickaite_2017}.

\section{Results}
\begin{figure}[t]
  \centering
  \includegraphics[width=1\columnwidth]{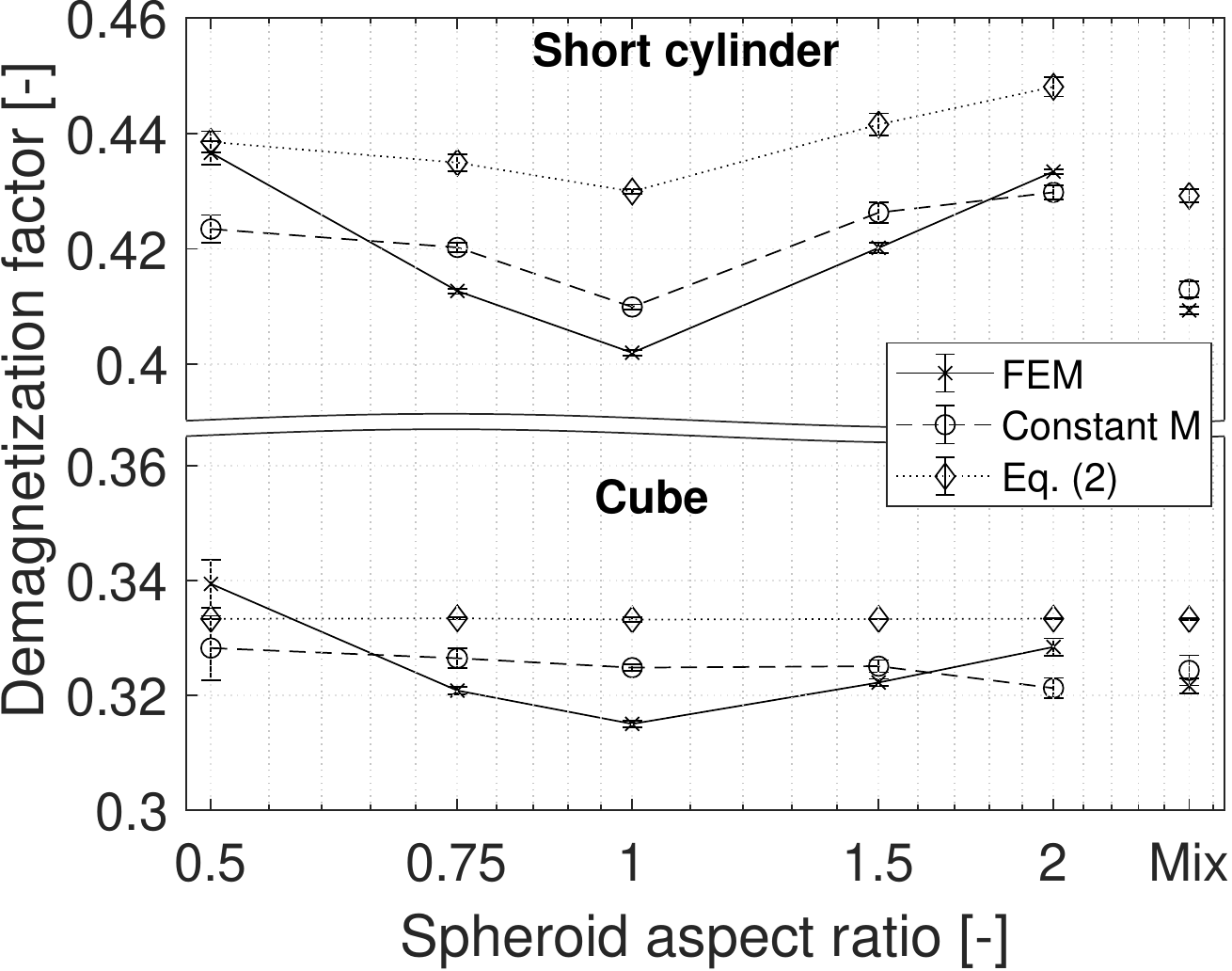}
  \caption{The demagnetization factor for a field along the $z$-direction, $N_\n{z}$, as function of the aspect ratio of the spheroidal particles for both the cube sample and the short cylindrical sample. The demagnetization factor is computed using two different numerical approaches, as well as Eq. (\ref{Eq.Demag}) which is appropriate for randomly packed spheres.}
    \label{Fig.AR_demag}
\end{figure}

The calculated average demagnetization factor, $N_\n{z}$, is shown in Fig. \ref{Fig.AR_demag} as function of the aspect ratio of the particles. The error bars indicate the standard deviation of the calculated demagnetization factor for five different samples with the same particle aspect ratio. As previously stated the demagnetization factor is computed using two different numerical frameworks. The demagnetization factor is also computed using the analytical expression for randomly packed spherical particles, Eq. (\ref{Eq.Demag}).

As can be seen from the figure, the overall value of the demagnetization predicted by Eq. (\ref{Eq.Demag}) is close, but the dependence on particle aspect ratio clearly seen in the FEM calculation is not observed. There is also a difference between the results predicted by FEM and those by the constant magnetization approach. The variation of Eq. (\ref{Eq.Demag}) with particle aspect ratio for the short cylinder is due to a slight variation of porosity with particle aspect ratio, as will be discussed subsequently. For the cube the sample demagnetization factor in Eq. (\ref{Eq.Demag}) is $D_\n{z}=0.33$, thus cancelling any dependence on porosity.

\begin{figure}[t]
  \centering
  \includegraphics[width=1\columnwidth]{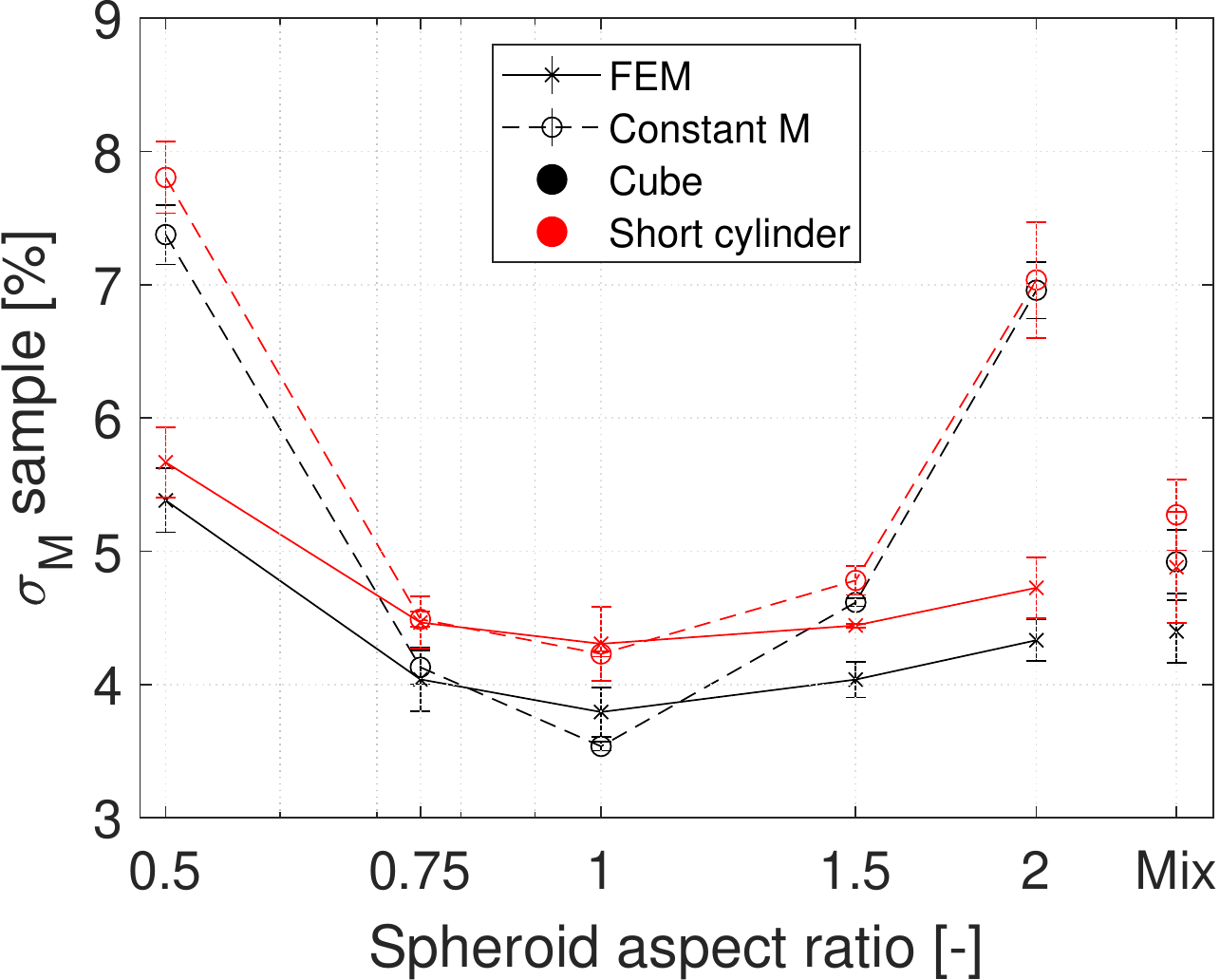}
  \caption{The relative standard deviation in magnetization across the sample as function of the aspect ratio of the spheroidal particles.}
    \label{Fig.AR_sigma}
\end{figure}

The FEM calculations also shows that there is an inhomogeneous magnetization both across the the powder sample and also within each individual particle. Shown in Fig. \ref{Fig.AR_sigma} is the relative standard deviation in internal magnetization across a given powder sample as function of the aspect ratio of the particles. In other words, the average magnetization in each particle is determined and Fig. \ref{Fig.AR_sigma} shows the average of this value for each sample, averaged for the five different samples considered from each packing. The error bar is the standard deviation across the five different samples at the same aspect ratio. The relative standard deviation is defined as
$\sigma_\n{M}=\frac{\sqrt{\langle{}(M-\langle{}M\rangle{})^2\rangle{}}}{\langle{}M\rangle{}}$. It is clearly seen that with particle shape different from a sphere, the spread in magnetization increases substantially.

\begin{figure}[t]
  \centering
  \includegraphics[width=1\columnwidth]{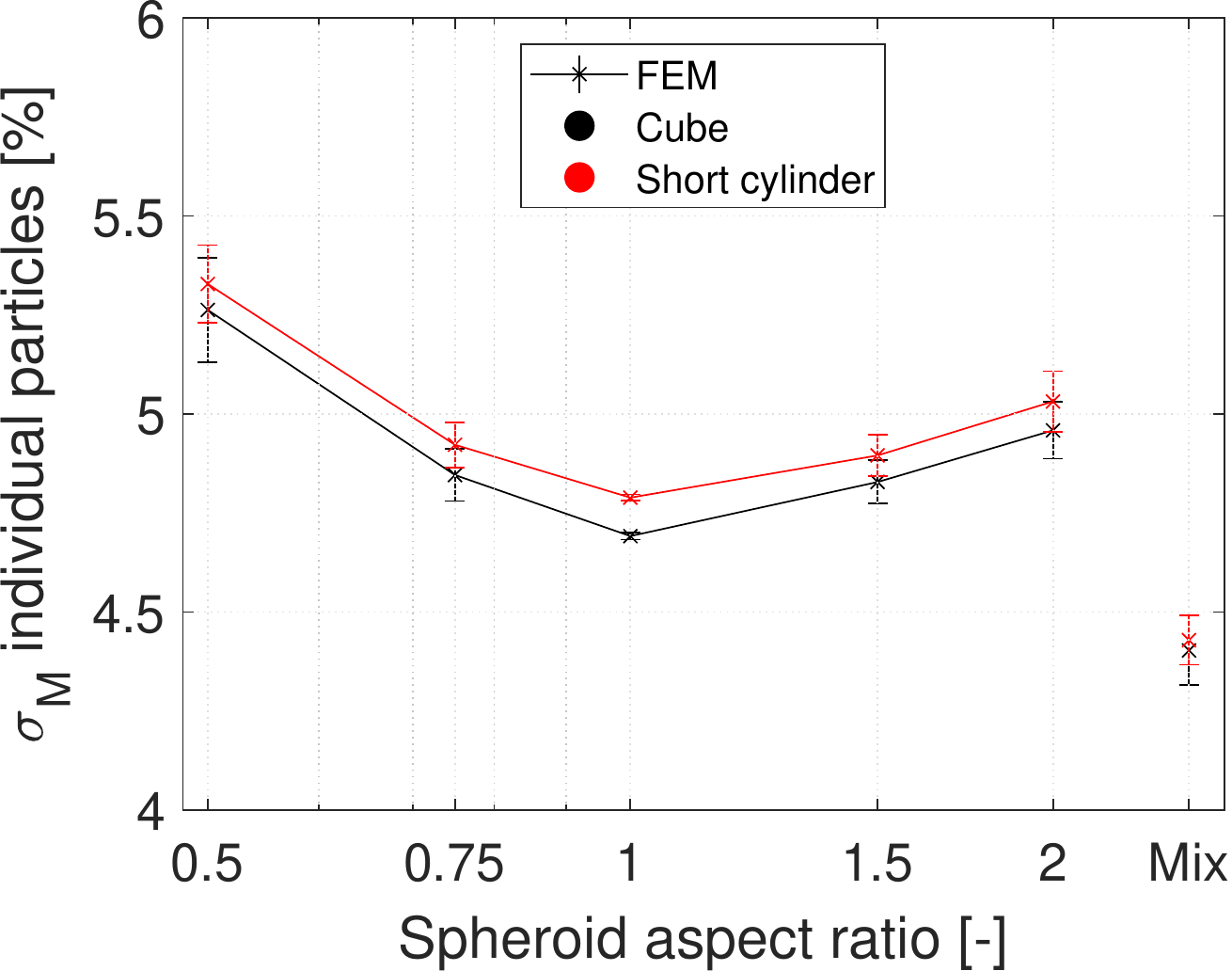}
  \caption{The relative standard deviation in magnetization within the individual particles as function of the aspect ratio of the spheroidal particles.}
    \label{Fig.AR_sigma_indi}
\end{figure}

In Fig. \ref{Fig.AR_sigma_indi} the average of the relative standard deviation in magnetization within each individual particles is shown for the FEM calculation, as function of the aspect ratio of the particles. Note that the demagnetization tensor model assumes that the magnetization is constant within each particle, and thus that model is not shown in this plot. The error bar is the standard deviation across the five different samples at the same aspect ratio. Compared to the relative standard deviation in the sample, the relative standard deviation within each particle is more constant around 5\%. The value of both the relative standard deviation of the sample and within the particles are very similar to the values reported for various particle size distributions for packed spheres in Ref. \cite{Bjoerk_2013}. The model with constant magnetization predicts a much larger spread in magnetization across the sample than the finite element model because the former is not able to account for the inhomogeneous magnetization within each individual particle. However, it remains clear from Fig. \ref{Fig.AR_demag} that Eq. (\ref{Eq.Demag}) can also be applied to correct experimentally measured magnetization data for spheroidal particles for demagnetization.

\subsection{Explaining the variation in $N_\n{z}$}

To explain the variation in demagnetization factor with the aspect ratio of the particles, we considered the packing of the spheroids in the sample. If the orientation of the particles is random, the packing is uniform and the demagnetization factor should not depend on the particle aspect ratio. The reason for this is that the demagnetization factors for the different sample faces must sum to 1 \cite{Moskowitz_1966} and if the sample is identical from all faces as for the cube, the demagnetization factors must also be same, and thus cannot depend on the particle aspect ratio.

That the demagnetization factor depends on the particle aspect ratio thus indicate that the powder sample as a whole is not uniform. We have investigated this by finding the orientation in space of the $c$-axis of all spheroids in the powder samples. The spherical coordinate system used is the ISO standard, with $\phi$ denoting the azimuthal angle and $\theta$ denoting the polar angle, as shown in Fig. \ref{Fig.Spheroid_illustration}.

\begin{figure}[t]
  \centering
  \includegraphics[width=1\columnwidth]{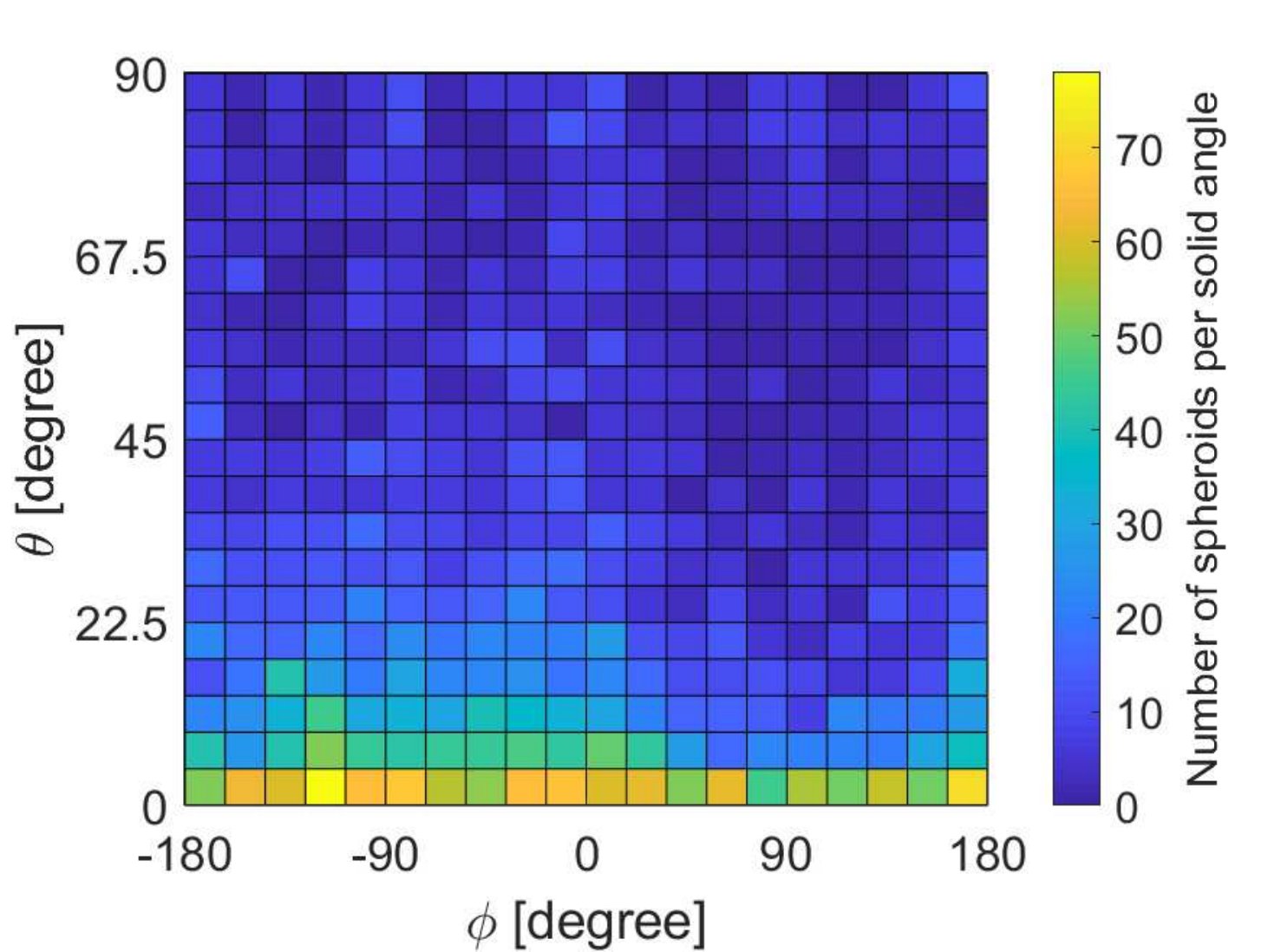}
  \caption{The distribution of the orientation of the $c$-axis for a cube packing with particles with an aspect ratio of 0.5.}
    \label{Fig.Angles_example}
\end{figure}

First an example of the orientation of the individual particles in a packing is shown in Fig. \ref{Fig.Angles_example}, where the angular orientation of the $c$-axis for all spheroidal particles in a cube packing is shown. The spheroids all have an aspect ratio of 0.5. As can be seen from the figure, there is a clear preference in the orientation of the particles with respect to the angle $\theta$.

\begin{figure}[t]
\begin{center}
\subfigure[]{}\includegraphics[width=0.47\textwidth]{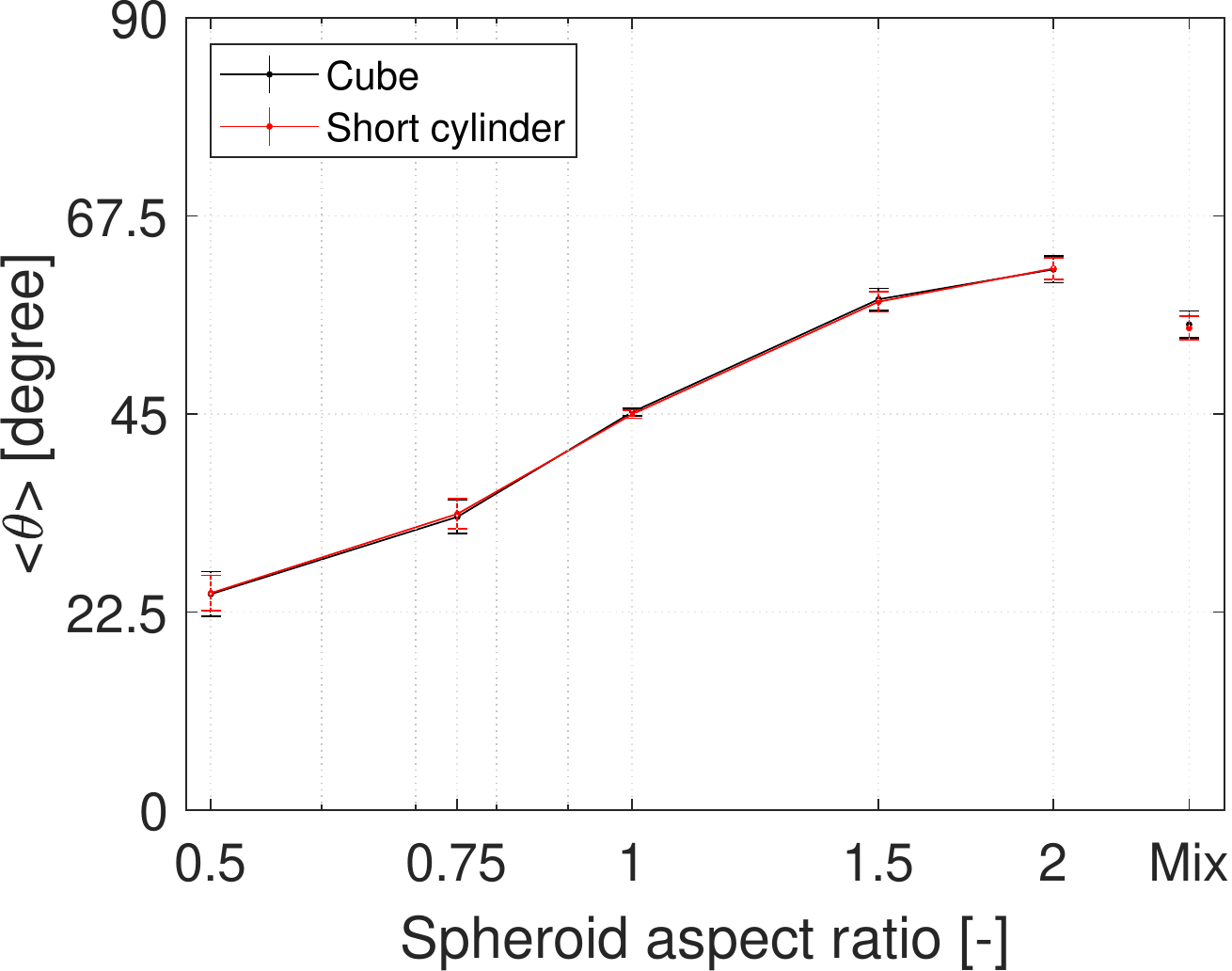}
\subfigure[]{}\includegraphics[width=0.47\textwidth]{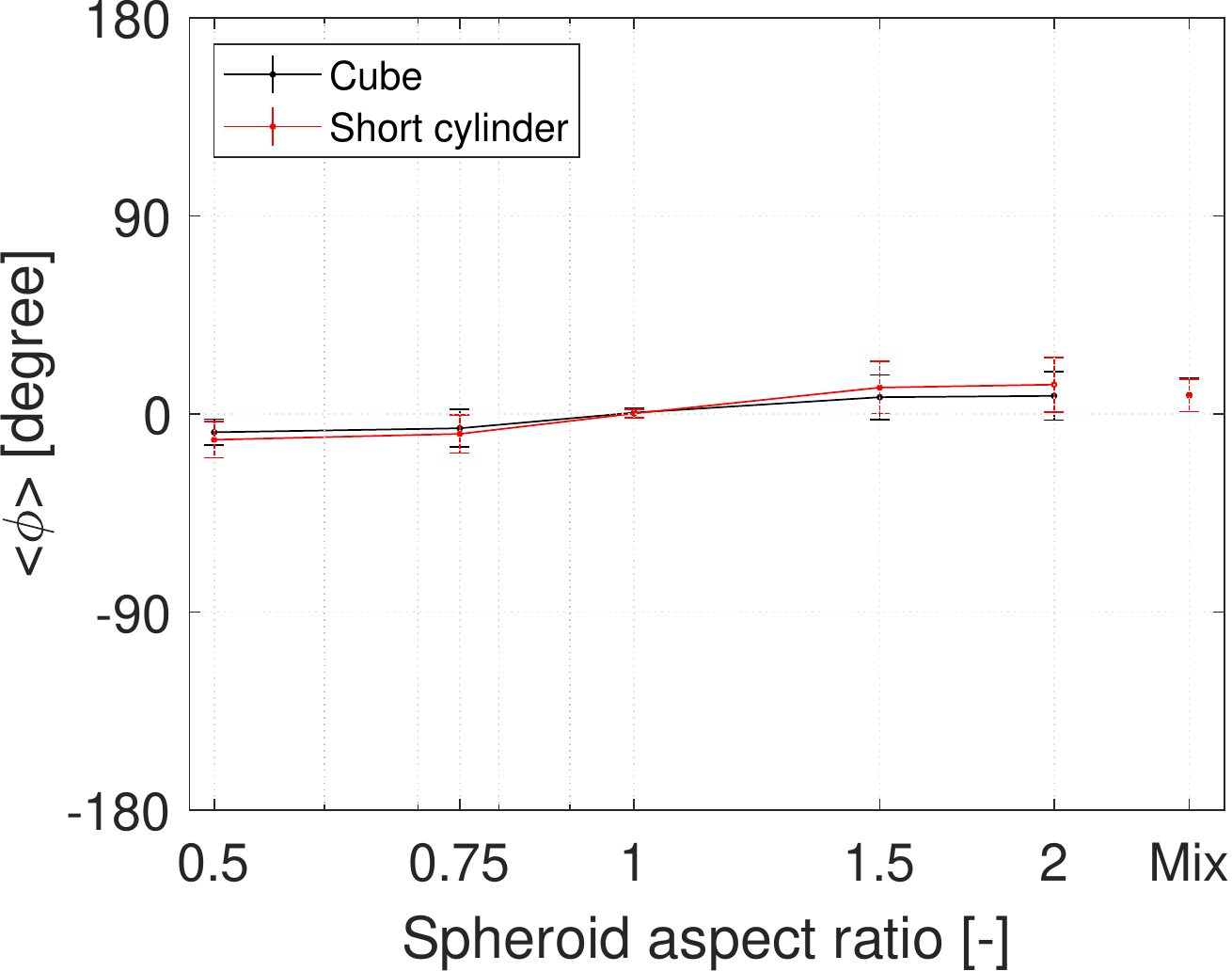}
\end{center}
\caption{The orientation of the spheroidal particles with respect to a) the spherical polar angle, $\theta$, and b) spherical azimuthal angle $\phi$, both as function of the aspect ratio of the spheroidal particles.}\label{Fig.theta_and_phi}
\end{figure}

For all packings the average orientation in the azimuthal direction, $\phi$, and the polar direction, $\theta$, as function of the particle aspect ratio is shown in Fig. \ref{Fig.theta_and_phi}. As can be seen from the figure, there is no dependence on $\phi$ but a strong dependence on $\theta$. The reason for this dependence on $\theta$ comes from the packing of the particles. As the first layer of spheroids is poured into the packing container these has a natural tendency to orientate relative to the flat bottom surface of the container, i.e. for an oblate to lay flat in the bottom surface with $\theta=0$. The following layers of spheroids poured into the sample container will also naturally orientate relative to the layers of the spheroids below them, i.e. for the oblate case having $\theta\approx{}0$.  It is this variation with aspect ratio that makes the powder sample as a whole non-uniform and that causes the demagnetization factor to depend on the particle aspect ratio.

The above results can be verified by experimental investigation using different approaches. The orientation of particles in a packed powder sample can be determined by X-ray tomography on a sample, while the demagnetization factor can be determined by measuring the magnetization of a powder sample in e.g. a vibrating sample magnetometer.

\subsection{Other factors influencing the demagnetization factor}
\begin{figure}[t]
  \centering
  \includegraphics[width=1\columnwidth]{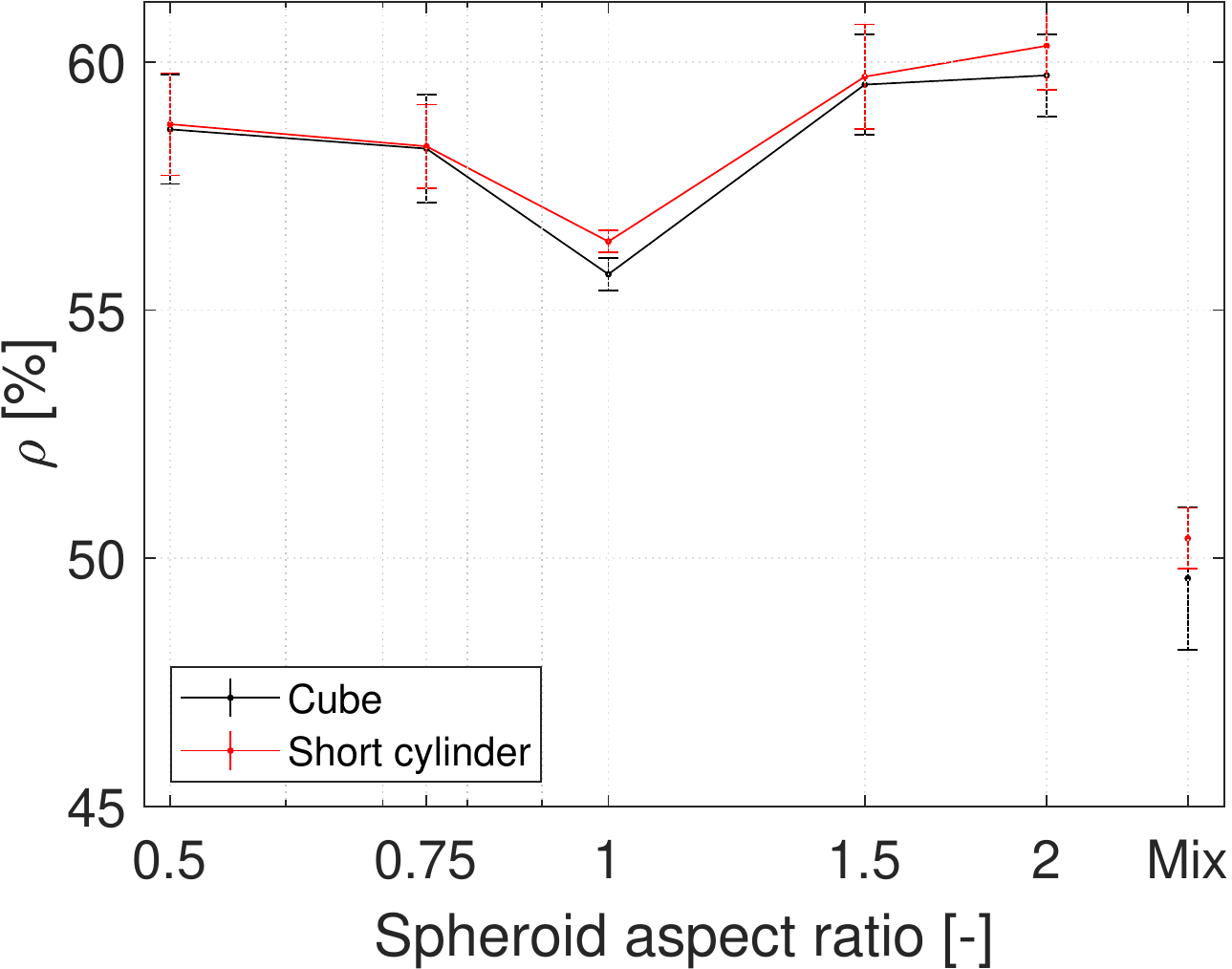}
  \caption{The relative density, $\rho$, as function of the particle aspect ratio.}
    \label{Fig.AR_rho}
\end{figure}

A number of other factors besides the particles orientation may affect the calculated demagnetization factor. These are the relative density (volume fraction of particles) and the overall sample shape. Shown in Fig. \ref{Fig.AR_rho} is the relative density as function of the particle aspect ratio. The relative density will influence the computed demagnetization factor as per Eq. (2). As can be seen from the figure there is a small variation of relative density with particle aspect ratio, but this variation does not show the same systematic variation as that of the demagnetization factor shown in Fig. \ref{Fig.AR_demag}. Therefore the slight variation of the relative density with particle aspect ratio does not cause the systematic variation showed previously.

\begin{figure}[t]
  \centering
  \includegraphics[width=1\columnwidth]{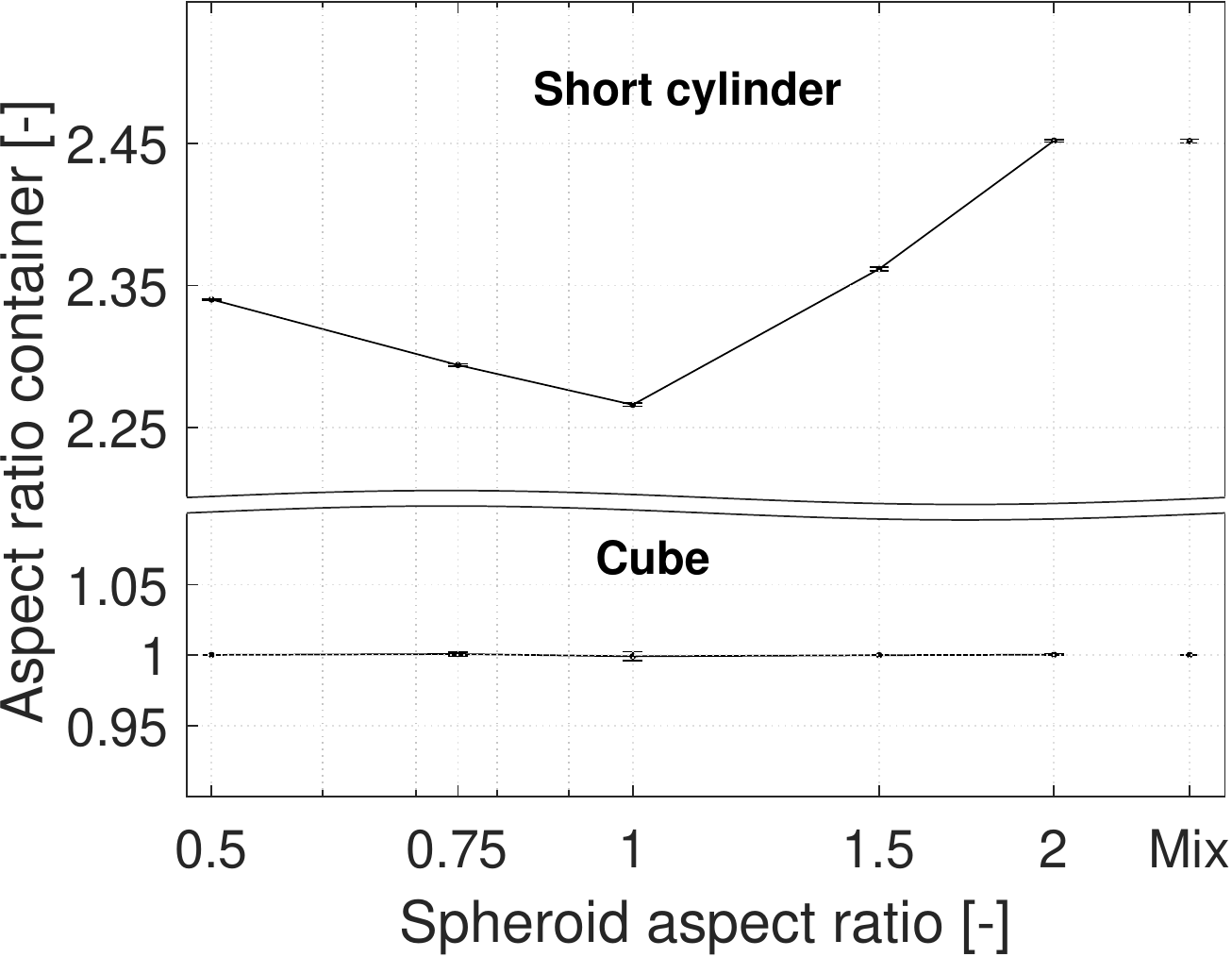}
  \caption{The aspect ratio of the sample, for all samples, as function of the particle aspect ratio.}
    \label{Fig.AR_AR_con}
\end{figure}

Another factor influencing the demagnetization factor is the overall shape of the powder sample. Although this has been chosen in the modeling setup, there may be a slight variation due to the packing of the powder, which could influence the computed demagnetization factor. The aspect ratio of the sample for both the cube and the short cylinder is shown in Fig. \ref{Fig.AR_AR_con} for all packings. As can be seen there is no variation of the aspect ratio for the cube, while there is a slight variation for the short cylinder. While the variation of the short cylinder may contribute slightly to the observed change in demagnetization factor for this geometry, the variation is not systematically the same as shown in Fig. \ref{Fig.AR_demag}, and thus it is at most a contributing factor in the variation of the demagnetization factor. And for the cube it is clear that the variation of the demagnetization factor is not caused by a variation of the samples' aspect ratio, as this is constant.

\section{Conclusion}
In conclusion, we have shown that the demagnetization factor for a powder sample of packed spheroid particles depend on the aspect ratio of the spheroids. This results from an non-uniform orientation of the spheroids within the powder sample when these are packed with gravity as the driving force. The orientation of the spheroid axes were shown to depend on the spheroid aspect ratio, due to the fact that the particle packing depended on the aspect ratio of the particles. The relative standard deviation of the magnetization in the powder was 3\%-8\%, increasing as the particle shape deviate from spherical.

\section*{Appendix}
\begin{figure}[t]
  \centering
  \includegraphics[width=1\columnwidth]{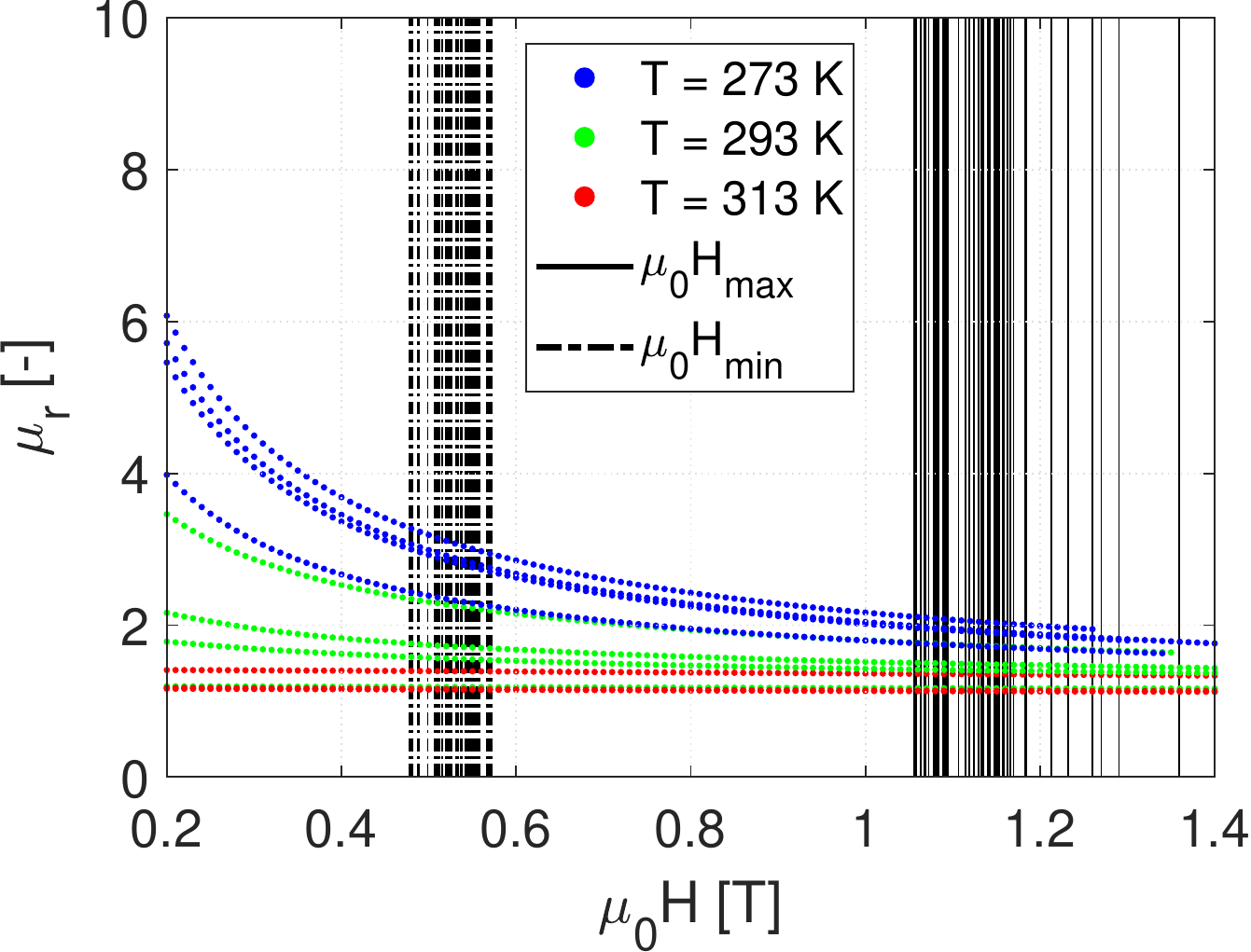}
  \caption{The relative permeability, $\mu_\n{r}$, at three different temperatures as function of the magnetic field for four ferromagnetic materials, Gd and LaFe$_\n{13-x-y}$Co$_\n{x}$Si$_\n{y}$ with (x=0.86, y=1.08), (x=0.94, y=1.01) and (x=0.97,y=1.07) respectively. Changing x and y results in a varying Curie temperature in the range 276-288 K. The data is taken from Ref. \cite{Bjoerk_2010d}. These materials cannot be individually identified on the figure. Also shown in the figure in the minimum and maximum magnetic field within the spheroidal particles modelled, for every individual sample.}
    \label{Fig.mu_r_material_data}
\end{figure}

A relative permeability of $\mu_\n{r}=2$ has been used throughout this work. The justification of using this value is given in Fig. \ref{Fig.mu_r_material_data}, which shows the relative permeability of four ferromagnetic materials at three different temperatures around room temperature. Shown on the same figure is also the minimum and maximum magnetic field within the spheroidal particles, for every individual sample, computed using $\mu_\n{r}=2$. It can clearly been seen that the four ferromagnetic materials shown on the figure have a relative permeability within a range of $\mu_\n{r}=1-3.5$ between the minimum and maximum value of the magnetic field. When this is considered together with the fact that small changes in the relative permeability does not change the demagnetization factor \cite{Bjoerk_2013}, this fully justifies the $\mu_\n{r}=2$ approximation. It should be mentioned that these four ferromagnetic materials are also magnetocaloric \cite{Smith_2012} and are used in room-temperature magnetic refrigeration devices \cite{Engelbrecht_2011,Balli_2012}.

\end{document}